# Are there rings around Pluto?

**J. J. Rawal [1], Bijan Nikouravan [2, 3]**

[1]The Indian Planetary Society (IPS) B-201, Lokmanya Tilak Road, Borivali (W), Mumbai – 400092, India
Email: ips.science@gmail.com

[2] Department of Physic, Islamic Azad University (IAU)-Varamin - Pishva Branch, Iran
[3]University of Malaya, 50603 Kuala Lumpur, Malaysia
Email: nikou@um.edu.my

**ABSTRACT**

Considering effects of tidal plus centrifugal stress acting on icy-rocks and the tensile strength thereof, icy-rocks being in the density range (1–2.4) g cm$^{-3}$ which had come into existence as collisional ejecta (debris) in the vicinity of Pluto at the time when Pluto-Charon system came into being as a result of a giant impact of a Kuiper Belt Object on the primordial Pluto, it is shown, here, that these rocks going around Pluto in its vicinity are under slow disruption generating a stable ring structure consisting of icy-rocks of diameters in the range (20–90) km, together with fine dust and particles disrupted off the rocks, and spread all over the regions in their respective Roche Zones, various Roche radii being in ~1/2 three-body mean motion resonance. Calculations of gravitational spheres of influence of Pluto which turns out to be 4.2 x 10$^6$ km for prograde orbits and 8.5 x 10$^6$ km for retrograde orbits together with the existence of Kuiper Belt in the vicinity of Pluto assure that there may exist a few rocks (satellites)/dust rings/sheets so far undiscovered moving in prograde orbits around the planet and few others which are distant ones and move around Pluto in the region between 4.2x10$^6$ km and 8.5x10$^6$ km in retrograde orbits.

**Key words**: Giant impact-Pluto-Charon System- Roche-limit- resonance- rings- new satellites- Kuiper-Belt.

## INTRODUCTION

Now that Pluto is no longer considered a planet and has been known to have three satellites, namely, Charon (Christy and Harrington, 1978, 1980; Harrington and Harrington, 1980; Harrington and Christy 1980a,b), Nix and Hydra (Weaver et al., 2006), a question has arisen as to whether Pluto has a ring structure and satellites so far undiscovered going around it. That NASA's robotic space-probe "New Horizons" has been on its way to Pluto, several authors in trying to answer the question, have analyzed the problem and would go on analyzing it untill the space- probe will reach Pluto in the year 2015. Some planetary scientists (Stern 1988, 1995, 2002; Stern et al. 1991, 1994, 1997a, b, 2003, 2006a, b; Steffl and Stern 2007; Steffl et al., 2006; Tholen and Buie 1988, 1990, 1997a, b) have come out with the prediction that Pluto may have a time-variable ring / dust sheets / partial rings around it which are far away from the planet. These rings were/are formed due to collisions of Kuiper Belt Objects with Nix and Hydra. Here, in this paper, we attempt to show that Pluto may have a stable ring structure and a few more satellites (rocks), yet undiscovered going around it.

## ROCHE LIMIT

The concept of Roche limit is well- known in the literature (Jeans 1928, 1960; Jeffery's, 1947). For a rigid body of density $\rho_s$ revolving around a primary having radius $R$ and density $\rho$, the Roche limit around the primary with respect to such a secondary is given by

$$R_{Roche} \approx 1.44 \left(\frac{\rho}{\rho_s}\right)^{1/3} R \qquad (1)$$

In what follows, we shall use system parameters as given in Table 1.

Table 1. Parameters for Pluto–Charon System that are used here.

| Object | Pluto | Charon | Nix | Hydra |
|---|---|---|---|---|
| Mass in kg | $1.4 \times 10^{22}$ | $0.14 \times 10^{22}$ | $5 \times 10^{-4}\, M_{Pluto}$ | $1 \times 0^{-5}\, M_{Pluto}$ |
| Radius in km | 1100 | 650 | 30 – 80 | 25 – 70 |
| Rotation period in days | 6.4 | 6.4 | —— | —— |
| Density in gm/c.c | 2.5* | 1.2 –1.22* | —— | —— |
| The mean distance from the parent body in (km) | $5.9 \times 10^9$ | 19,300 | 48,675 | 64,780 |
| Eccentricity of the orbit | 0.25 | May be small | —— | —— |
| Orbital period | 249 y | 6.4 d | ~ 25 d | ~ 38 d |

\* There is some uncertainty in the values of densities of Pluto and Charon.

## VARIOUS HYPOTHESES FOR FORMATION OF PLUTO-CHARON SYSTEM

After the discovery of Pluto's satellite Charon, several authors, (Mignard, 1981; Lin, 1981; McKinnon, 1984, 1988, 1989; Mckinnon and Muller 1988; Canup 2005), in trying to understand the formation of Pluto-Charon system, put forth several hypotheses such as fission and a giant impact origin. According to (Mignard 1981), there are numerous observed and theoretical facts which favour an origin of Charon by fission. Several authors have investigated the possibility for Pluto to be an ejected satellite of Neptune (Lyttleton 1936, 1953, 1961; Hoyle, 1975; Harrington and Van Flandern, 1979; and Dormand and Woolfson, 1980). After a so violent event the new angular speed of Pluto can give birth to a rotational instability which leads to the break-up of the primordial Pluto. If it was so, the radius of Pluto over that of Charon is to be close to 1.9. This result is in keeping with the new philosophy which is currently emerging in the field of planetology. The solar system is definitely more diversified than it was thought. Then, there is no reason to believe in a unique process to originate satellites, and mechanisms as varied as accretion, capture, fission and giant impact origin have likely been efficient throughout the Solar System.

Lin (1981) proposed that Pluto-Charon System might have been formed by binary fission of a rapidly rotating body. If, as a result of fission, Charon was formed just inside its own tidal radius, the mass ratio of Charon to Pluto must not exceed 0.25. Otherwise the resultant spin angular momentum of Pluto would cause it to break up again. The mass ratios must be greater than 0.05 in order for Charon to form outside the unstable co-rotation radius and subsequently evolve to the present stable co-rotation radius rather than be driven back to Pluto. The observational value for mass ratios is 0.1, which is consistent with the limit set by his own hypothesis (This was the case in 1981 when the correct mass ratio of Charon to Pluto and other Pluto-Charon System parameters were exactly unknown, even today also some system parameters are not known exactly). Furthermore, it is very close to the critical value for mass ratio in which the initial object has enough angular momentum to become secularly unstable. Lin (1981) felt that further observations on the mass, size, binary separation and the density of Pluto and Charon would provide new insight into the process of binary fission and planetary formation (also see Foust et al., 1997).

Precise determination of diameter of Pluto and Charon along with the total mass of the system provide a powerful basis for constraining the origin of this enigmatic planetary pair. The work of Mckinnon (1989) focuses on the angular momentum budget of Pluto-Charon, taking as the point of departure from earlier work of Lin (1981). Because of the large angular momentum density of the system, he argued for an impact or collisional origin. He then addressed aspects of the required impact process and compared them with somewhat similar hypotheses for the Moon's origin. It has been recognized for some time that Pluto-Charon's J value is high, but it was not known to be very high as it has been found to be now. Lin (1981) and Mignard (1981), therefore, advocated fission of a single original object. McKinnon (1984), Burns (1986) and Peale (1986) suggested a large-body impact. For comparison, the Earth-Moon system has a J of 0.115, and even this has been judged great enough for impact over spinning to be seriously considered [Durisen and Gingard, (1986)]. The logical cause of Pluto-Charon's large J is a large-body impact. An impact origin is physically plausible as it is for suspected binary asteroids (Weidenschilling et al., 1989, Weidenschilling 2002; see also Mckinnon and Muller, 1988).

Canup (2005) used hydro dynamical simulations to demonstrate that the formation of Pluto-Charon by means of a large collision is quite plausible. He also observed that such an impact probably produced an intact Charon, although it is possible that a disc of material orbited Pluto from which Charon later accumulated. These findings suggested that collisions between 1000-kilometer-class objects occurred in the early inner Kuiper Belt.

## ROLE OF A GIANT IMPACT ORIGIN HYPOTHESIS, ROCHE LIMIT AND A THREE-BODY MEAN MOTION RESONANCE IN THE FORMATION OF A STABLE RING STRUCTURE AROUND PLUTO

According to a Giant Impact Origin Hypothesis, Pluto-Charon System came into being as a result of a collision of a big Kuiper Belt Object (1000 km size) with primordial Pluto. If this is so, it is natural to expect collisional ejecta (debris, fragments) spread all around in the Pluto-Charon System and revolve around Pluto-the biggest of all remnants of the catastrophic collisional event. This shows that the region in the Pluto-Charon System may not be clean but would be full of small or big collisional ejecta (debris, fragments in the form of rocks) and also the facts that the two newly discovered satellites Nix and Hydra of Pluto (Weaver et al.2006) are in proximity to Pluto and Charon, they are on near-circular orbits in the same plane as Pluto's large satellite Charon, along with their apparent locations in or near high order mean motion resonances, all probably result from their being constructed from collisional ejecta that originated from the Pluto-Charon System formation event (Stern et al., (2006a, b). Stern et al., (2006a, b) also argue that dust rings of variable optical lengths form sporadically in the Pluto System far away from Pluto due to collisions of Kuiper Belt objects with Nix and Hydra and that rich satellite systems may be found perhaps frequently-around other large Kuiper belt objects. Let us, therefore, consider rocks (satellites) having densities in the range (1–2.4) g cm$^{-3}$ (Table 2) lying in the neighborhood of Pluto. We now calculate the Roche limits, $R_i$, with respect to rocks of densities in the range (1–2.4) g cm$^{-3}$. These are shown in (Table 2). Taking tensile strength, s, for an ice-rock to be $3\times10^6$ dyne cm$^{-2}$ (Jeffreys, 1947) in the formula,

$$r = \left(\frac{19a^3s}{32GM\rho_s}\right)^{1/2} \quad (2)$$

Where $r$ is the reduced radius of a satellite upto which the satellite has been ruptured; $a$, the radial distance at which the satellite (rock) is ruptured to its maximum; $G$, the Universal gravitational constant; M, the mass of Pluto; $\rho_s$, the density of satellite and $s$, is the tensile strength. We calculate the reduced radius up to which each rock has been ruptured at various Roche radii $R_i$, rock density $\rho_{si}$ being in the range (1–2.4) g cm$^{-3}$. This is shown in Table 2.

**Table 2.** Location of Stable Ring Structure around Pluto. Table showing Roche radial distances, minimum radii that satellites can retain at corresponding $R_i$ and when they graze the planet; revolution periods, $T_i$, corresponding to $R_i$. All these parameters corresponding to densities, $\rho_{si}$, in the range (1―2.4) g cm$^{-3}$

| Densities, $\rho_{si}$ of rocks(secondary) in (g cm$^{-3}$) | Corresponding Roche limit $R_i$ (km) | Minimum radius r (km) that a satellite can retain at corresponding $R_i$ | Minimum radius r (km) That a satellite can retain when it grazes the planet. | $T_i$ [revolution period in days at corresponding Roche radial distance ($R_i$)] |
|---|---|---|---|---|
| $\rho_{s\,8}$ = 1     | $R_8$ = 2149.5 | $r_8$ = 43.55 | $r_8$ = 16   | $T_8$ = 0.2369 |
| $\rho_{s\,7}$ = 1.2   | $R_7$ = 2022.1 | $r_7$ = 36.28 | $r_7$ = 14.6 | $T_7$ = 0.2162 |
| $\rho_{s\,6}$ = 1.4   | $R_6$ = 1920.9 | $r_6$ = 31.28 | $r_6$ = 13.5 | $T_6$ = 0.2001 |
| $\rho_{s\,5}$ = 1.6   | $R_5$ = 1837.2 | $r_5$ = 27.21 | $r_5$ = 12.6 | $T_5$ = 0.1872 |
| $\rho_{s\,4}$ = 1.8   | $R_4$ = 1767.7 | $r_4$ = 24.21 | $r_4$ = 11.9 | $T_4$ = 0.1766 |
| $\rho_{s\,3}$ = 2.0   | $R_3$ = 1706.0 | $r_3$ = 21.72 | $r_3$ = 11.3 | $T_3$ = 0.1675 |
| $\rho_{s\,2}$ = 2.2   | $R_2$ = 1653.5 | $r_2$ = 19.81 | $r_2$ = 11.0 | $T_2$ = 0.1598 |
| $\rho_{s\,1}$ = 2.4   | $R_1$ = 1604.1 | $r_1$ = 18.12 | $r_1$ = 10.3 | $T_1$ = 0.1528 |

From Table 2, it is clear that the reduced radii up to which the rocks have been ruptured are in the range 15 to 45 km with respect to Roche radii, and are in the range 10 to 16 km at the radial distance grazing the planet. All these rocks lie within the radial distance ~2,500 km around Pluto. This, we believe, form a ring around Pluto. As tidal disruption goes on in this region, the region is full of fine dust, particles and small or big pebbles and rocks, forming a stable ring structure around Pluto. Resonance theory states that if $n_1, n_2, n_3, (n_i = \frac{2\pi}{T}, n_1 > n_2 > n_3)$, are mean motions of three secondaries in circular orbits, then condition for frequent occurrence of mirror configuration (Dermott 1968 a, b, 1973; Greenberg, 1973; Gold reich, 1965 a, b; Ovenden et al., 1974; Roy and Ovenden, 1955; Rawal 1981, 1989; Goldreich and Soter, 1966; and references given therein) is given by the Equation

$$\alpha n_1 - (\alpha + \beta)n_2 + \beta n_3 = 0 \quad (3)$$

Where α, β are small and mutually prime positive integers. It follows from Equation (3) that in a reference frame rotating with the mean motion of any one of the three secondaries, the relative mean motions $n'_i$ of other two are commensurate, and that in a frame I (that of the innermost secondary), we have

$$\frac{n'_2}{n'_3} = \left(\frac{n_2-n_1}{n_3-n_1}\right) = \left(\frac{\beta}{\beta+\alpha}\right) \quad (4)$$

In terms of revolution periods, Equation (4) is written as

$$\frac{n'_2}{n'_3} = \frac{T_3(T_2-T_1)}{T_2(T_3-T_1)} = \left(\frac{\beta}{\beta+\alpha}\right) \quad (5)$$

In order to know whether a triad of successive secondaries at various Roche radial distances $R_i$ (given in Table 2) is in stable three body mean motion resonance, we calculate revolution periods $T_i$ (shown in Table 2) corresponding to various $R_i$ to find corresponding mean motions. From Equation (6) we find that a triad of successive secondaries at various $R_i$ has relative mean motion ratio ~1/2 throughout, that is, they follow three-body mean motion resonance relation given by

$$n_1 - 2n_2 + n_3 = 0 \quad (6)$$

This shows that various rocks (satellites) at or near these resonance orbits are in reasonably stable resonant orbits, and if they existed there, then they still exist there. In other words, rocks and particles along with fine dust which got disrupted off the rocks form a stable ring structure in the vicinity of Pluto which is within the distance ~2500 km from the centre of the planet.

**GRAVITATIONAL SPHERE OF INFLUENCE OF PLUTO FOR PROGRADE AND RETROGRADE ORBITS AND THEIR SIGNIFICANCE FOR EXISTENCE OF UNKNOWN SATELLITES IN THE SYSTEM.**

In order to know the existence of distant unknown satellites in the Pluto- Charon System, we would like to know how far the gravitational influence of Pluto is, that is, how large the gravitational sphere of influence of Pluto is. We, therefore, calculate here, the boundaries to the gravitational sphere of influence of Pluto for prograde and retrograde orbits using King-Innanen formula. Innanen (1979) modified the King (1962) formula for putting in the equation for acceleration in a revolving co-ordinate frame with an additional Coriolis term of magnitude, $2\Omega v_r$ where $v_r$ is the velocity of the secondary relative to the primary. Here in our case, we consider Moon-Pluto-Sun System, and therefore, the primary is the Sun and the secondary is Pluto. The familiar right hand rule immediately shows that the Coriolis term is always directed radially between the secondary (Pluto) and the primary (Sun). It counteracts the primary's gravity for the direct motion of the moon, but effectively supplements the primary's gravity for retrograde motion. For the limiting direct and retrograde radii, $r_d$ and $r_r$ respectively, of a moon around a planet (here Pluto) in the most general case where the planet (here Pluto)'s orbit has eccentricity e and the pericentric distance $R = R_p = a(1-e)$, a, being the mean distance of the planet (Pluto), we have

$$\frac{r_r}{r_d} = [f(e)]^{2/3} \tag{7}$$

Where

$$r_d = \left[\frac{1}{(f(e))^2}\frac{m}{M}\right]^{\frac{1}{3}} R_p \tag{8}$$

$$f(e) = \left[\frac{5+e+2(4+e)^{1/2}}{3+e}\right] \tag{9}$$

Here, $m$, is the mass of the planet (here Pluto), M, is the mass of the Sun. Therefore, calculation for the gravitational sphere of influence of Pluto for prograde orbits turns out to be $4.2 \times 10^6$ km and for retrograde orbits, it is $8.5 \times 10^6$ km. The distances of Charon, Nix, Hydra are 19300, 48675, 64780 km respectively. These satellites are very much inside the boundary of the gravitational sphere of influence for prograde satellites. The boundary of the gravitational sphere of influence for retrograde satellites is far far away.

At the time of a giant impact, collisional ejecta (debris) were likely to be thrown far away. It is, therefore, likely that there may be few prograde satellites (rocks) revolving around Pluto, between Pluto and Charon and beyond Hydra within the distance $4.2 \times 10^6$ km, and a few retrograde satellites (rocks) revolving around the planet between the distance $4.2 \times 10^6$ km and $8.5 \times 10^6$ km. Moreover, in the vicinity of Pluto, there is Kuiper belt. It is likely that some Kuiper belt Objects might have been captured by Pluto making them its satellites and / or injected into the gravitational sphere of influence of Pluto thereby becoming its satellites. They may be prograde moving if they had entered into the gravitational sphere of influence of Pluto for prograde satellites, that is, within the distance $4.2 \times 10^6$ km, and may be retrograde moving if they are outside it, but inside the gravitational sphere of influence of Pluto for retrograde motion, that is, distance between $4.2 \times 10^6$ km and $8.5 \times 10^6$ km. This analysis shows that Pluto may have a few, yet undiscovered satellites going around it, a few of them even moving in retrograde direction.

**CONCLUSION**

Here, it is shown that a stable ring system consisting of small rocks having densities in the range (1–2.4) g cm$^{-3}$ and diameters in the range (20–90) km along with fine dust and particles disrupted off these rocks, may exist around Pluto within the distance ~2500 km from the centre of the planet. There may also exist a few satellites (rocks) other than already known. If these satellites (rocks) orbit the planet within the distance $4.2 \times 10^6$ km, then they are protrude moving, and if they orbit the planet in the region between $4.2 \times 10^6$ and $8.5 \times 10^6$ km, then they are retrograde moving. These satellites may be collisional ejecta (debris) which came into existence as a result of catastrophic collision event which formed Pluto-Charon System or rocks captured by Pluto from Kuiper belt and made them its own satellites or Kuiper belt Objects injected into Pluto's gravitational sphere of influence by some gravitational perturbations due to Neptune or Oort's cometary cloud as a whole becoming its satellites

**ACKNOWLEDGEMENTS**

Thanks are due to Professors K. Sivaprasad and S. Ramadurai (both retired from Tata Institute of Fundamental Research, Mumbai) for helpful discussions and useful suggestions and Mr. Hemal S. Shah, Mr. D. K. Sahani and Mr. Darshan K. Vishwakarma for their technical assistance.

Thanks are also due to anonymous referees for their constructive criticisms and helpful suggestions that have greatly improved the content of this paper.